\documentclass[aps,pre,superscriptaddress,showpacs,reprint]{revtex4-1}
\usepackage[utf8]{inputenc}
\usepackage{amsmath,amssymb,graphicx}
\usepackage[dvipsnames]{xcolor}
\usepackage[colorlinks=true,citecolor=Green]{hyperref}
\newcommand{\fig}[2]{\includegraphics[scale = #1]{#2}}
\newcommand{\ie}{\emph{i.e.}}
\newcommand{\eg}{\emph{e.g.}}

\definecolor{r_dark}{RGB}{233, 71, 73}
\definecolor{r_middle}{RGB}{241, 140, 141}
\definecolor{r_pale}{RGB}{249, 209, 209}

\begin{document}
\title{Model-based evaluation of scientific impact indicators}
\author{Matúš Medo}
\email{matus.medo@unifr.ch}
\affiliation{Physics Department, University of Fribourg, CH-1700 Fribourg, Switzerland}
\author{Giulio Cimini}
\email{giulio.cimini@imtlucca.it}
\affiliation{IMT School for Advanced Studies, 55100 Lucca, Italy}
\affiliation{Istituto dei Sistemi Complessi (ISC)-CNR, 00185 Rome, Italy}
\date{\today}

\begin{abstract}
Using bibliometric data artificially generated through a model of citation dynamics calibrated on empirical data, we compare several indicators for the scientific impact of individual researchers. 
The use of such a controlled setup has the advantage of avoiding the biases present in real databases, and allows us to assess which aspects of the model dynamics and which traits of individual researchers 
a particular indicator actually reflects. We find that the simple citation average performs well in capturing the intrinsic scientific ability of researchers, whatever the length of their career. 
On the other hand, when productivity complements ability in the evaluation process, the notorious $h$ and $g$ indices reveal their potential, yet their normalized variants do not always yield a fair comparison 
between researchers at different career stages. Notably, the use of logarithmic units for citation counts allows us to build simple indicators with performance equal to that of $h$ and $g$. 
Our analysis may provide useful hints for a proper use of bibliometric indicators. Additionally, our framework can be extended by including other aspects 
of the scientific production process and citation dynamics, with the potential to become a standard tool for the assessment of impact metrics.
\end{abstract}

\maketitle

\section{Introduction}
The quantitative study of the productive and communication aspects of science, known as {\em Scientometrics}, is nowadays well established. 
This discipline focuses mainly on the analysis of citation statistics of the academic literature in order to identify suitable indicators for the impact of research~\cite{Mingers2015}. 
Well-known and widely used examples of impact indicators include the journal impact factor~\cite{Garfield1972} and the $h$-index~\cite{Hirsch2005}, but several (more than one hundred~\cite{Wildgaard2014}) 
alternatives have been proposed---see~\cite{Alonso2009,Egghe2010,Waltman_rev} for recent reviews of the field. Importantly, these metrics are intended to measure scientific impact, and not quality or importance. 
Yet, nowadays they play a central role in the measurement and evaluation of research performance (at the level of individual researchers, research groups and institutions), 
despite the numerous warnings from the scientific community~\cite{Lawrence2008,Hicks2015,Werner2015}. The issue is critical especially at the level of individual researchers, 
as it can affect received funds and grants---not to mention employment and career.

Recently, Wildgaard and colleagues~\cite{Wildgaard2014} pointed out the need to examine author-level indicators in relation to what they are supposed to reflect 
and especially to their specific limitations. Indeed, by comparing the key concepts of several metrics, they showed that no indicator alone can capture the overall impact of a researcher, 
which is instead better characterized by a combination of indexes. Such a combination is however not unique, and depends on the particular type of assessment to be made. 
Evaluation of impact indicators is also complicated by the availability and reliability of the bibliometric databases (such as {\em Web of Science}, {\em Scopus}, {\em Google Scholar} 
and {\em Microsoft Academic Search})~\cite{Waltman_rev}. In fact, these databases suffer to various extent from the lack of quality control~\cite{Jacso2006} and partial coverage. 
The latter problem is relevant especially in the fields of social sciences and humanities~\citep{Sivertsen2012}, which may have a strong national or even regional orientation 
and thus target local journals and books~\citep{Nederhof2006}, and for computer science and engineering---where conference proceedings play an important role, but they are ofter not counted or counted twice 
(as the work is published both as proceedings and as regular journal paper). All these facts cause the measured impact of a researcher to depend on the specific data used in the calculation~\cite{Wildgaard2014}. 
Besides, these data are polluted by improper citation practices used by researchers (like boosting self or friend's citations, or satisfying referees) that are not related at all to the acknowledgment 
of a paper's importance~\cite{Werner2015}. 

\begin{figure*}
\centering
\fig{0.7}{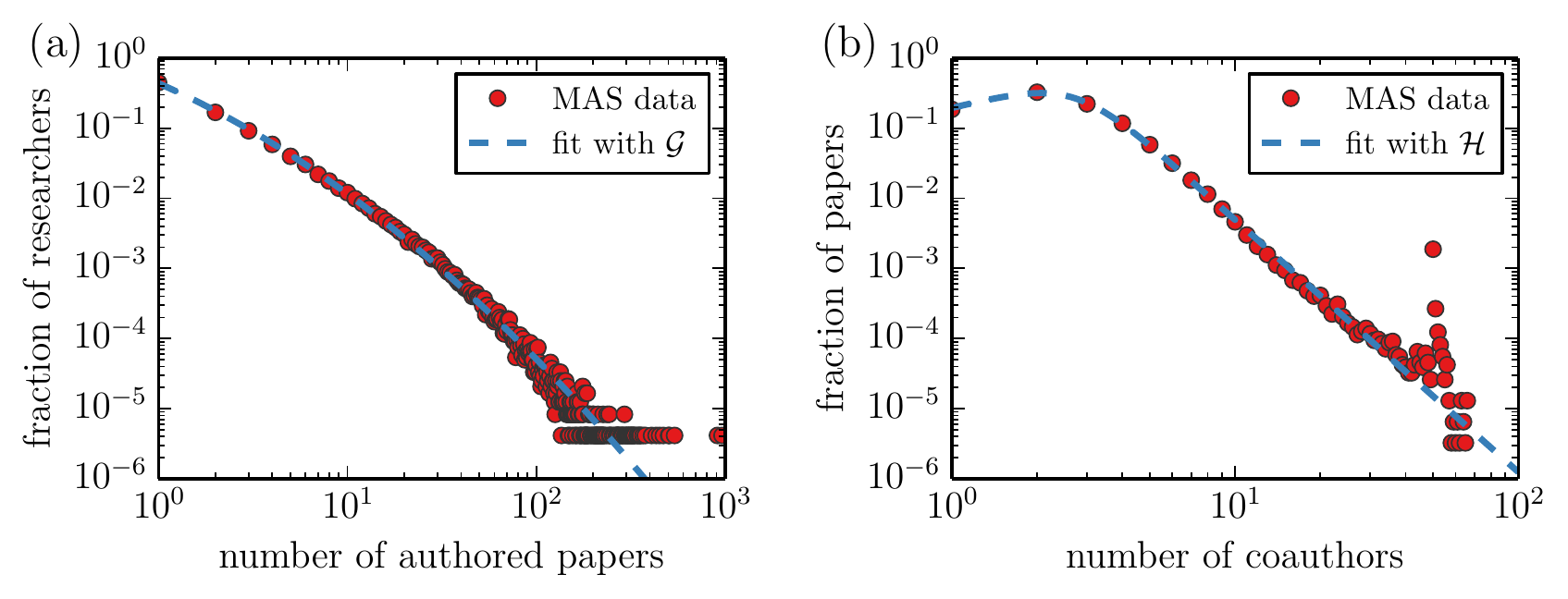}
\caption{Functional fits of the MAS data presented in the text: (a) the number of publications for researchers, and (b) the number of co-authors for papers. The fit functions are 
$\mathcal{G}(k)=3.48/\exp[-(\ln k+3.5)^2/5.9]$ and $\mathcal{H}(d)=19.7d/(100+d^{4.6})$, respectively.
In the log-log scale, the coefficients of determination ($R^2$) are 0.92 and 0.89, respectively. For the co-author distribution, 
the peak at $d=50$ co-authors is due to large-scale collaborations in particle physics and astrophysics, whose members are only partially covered in our data.\\
The underlying Microsoft Academic Search (MAS) data that we present here were collected using the API of the service to obtain unique IDs for the authors of papers 
published by the American Physical Society (APS) in years 1893--2009; this was successful for 71\% of the APS papers. Excluding self-citations, 
the resulting data comprise 2,427,367 citations among 326,586 papers authored by 244,538 researchers. Thanks to having unique author IDs, 
the use of MAS data avoids the common name disambiguation problem in bibliometric data~\cite{schulz2014exploiting} which is vital for the analysis of co-authorship patterns~\cite{Sarigol2014}.}
\label{fig:empirical}
\end{figure*}

On the theoretical side, the scientific community has devoted much effort to unveil the dynamics of the citation process, the main focus being that of explaining the extremely skewed distribution 
of the number of citations received by scientific papers (see for instance~\cite{Redner2005}). Notably, in 1976 Price~\cite{Price1976} was the first to tackle this issue by using a model 
based on preferential attachment, a process for which some quantity associated to the entities of a system (the number of citations of scientific publications, in our context) 
is distributed and grows according to how much these entities already have. Later, this model has been much studied and generalized (see~\cite{Albert2002} for a review of the field). 
Importantly, the original version of the model predicts a strong relation between a paper's age and its citation count, but significant deviations from this behavior are found in bibliometric data~\cite{Newman2009}. 
It has been recently pointed out that to model citation dynamics well, preferential attachment has to be combined 
with intrinsic paper {\em relevance}: a heterogeneously distributed ``quality'' (fitness) that decays with time~\cite{Medo2011,Eom2011,Wang2013}. 
These models are then capable of generating artificial data that closely resemble real citation networks~\cite{Medo2014}. 

Building on this modeling framework, we aim to perform a comparative evaluation of various scientific impact indicators, in the same spirit of~\cite{Wildgaard2014} but on a quantitative basis provided 
by the use of an artificial setting. In particular, we extend a previous model constrained on the citation dynamics of scientific papers~\cite{Medo2011} by assuming that researchers are endowed with intrinsic 
productivity and ability levels---the latter determining the fitness of their authored papers, that in turn make connections to the existing body of literature according to the 
modified preferential attachment mechanism described above. The artificial bibliometric data generated by the model then allow us to compute a variety of impact indicators, which can be compared 
with the individual traits of researchers in order to determine what these indicators actually capture. We can thus identify the indicators which properly rank authors, and those that fail in this task. 
Notably, our controlled and simplified setup has the advantage to generate citation records which are free from the biases present in real databases that can hinder this kind of analyses.

The paper is organized as follows. Section II describes the model used to generate citation data, and Section III provides the definitions of the impact indicators that we compare and evaluate. 
Results of the analysis are reported in Section IV, while Section V concludes the work and outlines the possibilities for further improving the artificial framework 
by including additional relevant aspects of the citation dynamics, such as differences between scientific disciplines~\cite{Radicchi2008} and journal reputation~\cite{judge2007causes} 
(see~\cite{scharnhorst2012models,watts2014simulating} for recent progresses in modeling the various aspects of the research and citation process).

\section{Model and Artificial Data}
\label{sec:model}
In this work we use a model which extends the one suggested in~\cite{Medo2011}. The system is composed of researchers (or authors) and papers, indicated by Latin and Greek letters respectively. 
Time runs in discrete time steps corresponding to months, and the simulation spans over $T$ months. There are $A$ authors in the system. For the sake of simplicity, we assume that their number is fixed 
and that they are all active during the whole simulation, but this assumption is relaxed later on. Each author $i$ is characterized by ability $a_i$ and productivity $k_i$ 
(\ie, the total number of papers that $i$ will co-author). In line with the exponential distribution of total paper relevance presented in~\cite{Medo2011}, author ability is drawn 
from the exponential distribution $\mathcal{F}(a) = a_0 + \mathrm{e}^{-a}$. The parameter $a_0$, acting as the minimal author ability, is motivated by the presence of some ``entrance barriers'' in academia
which guarantee that all authors have some minimal ability value and thus their papers have some minimal level of relevance to the community. Author productivity is drawn from 
$\mathcal{G}(k)=3.48/\exp[-(\ln k+3.5)^2/5.9]$ which has been obtained by fitting the real distribution of authored papers in the Microsoft Academic Search (MAS) data; see Figure~\ref{fig:empirical}(a) 
and the description therein.

\begin{figure*}
\centering
\fig{0.7}{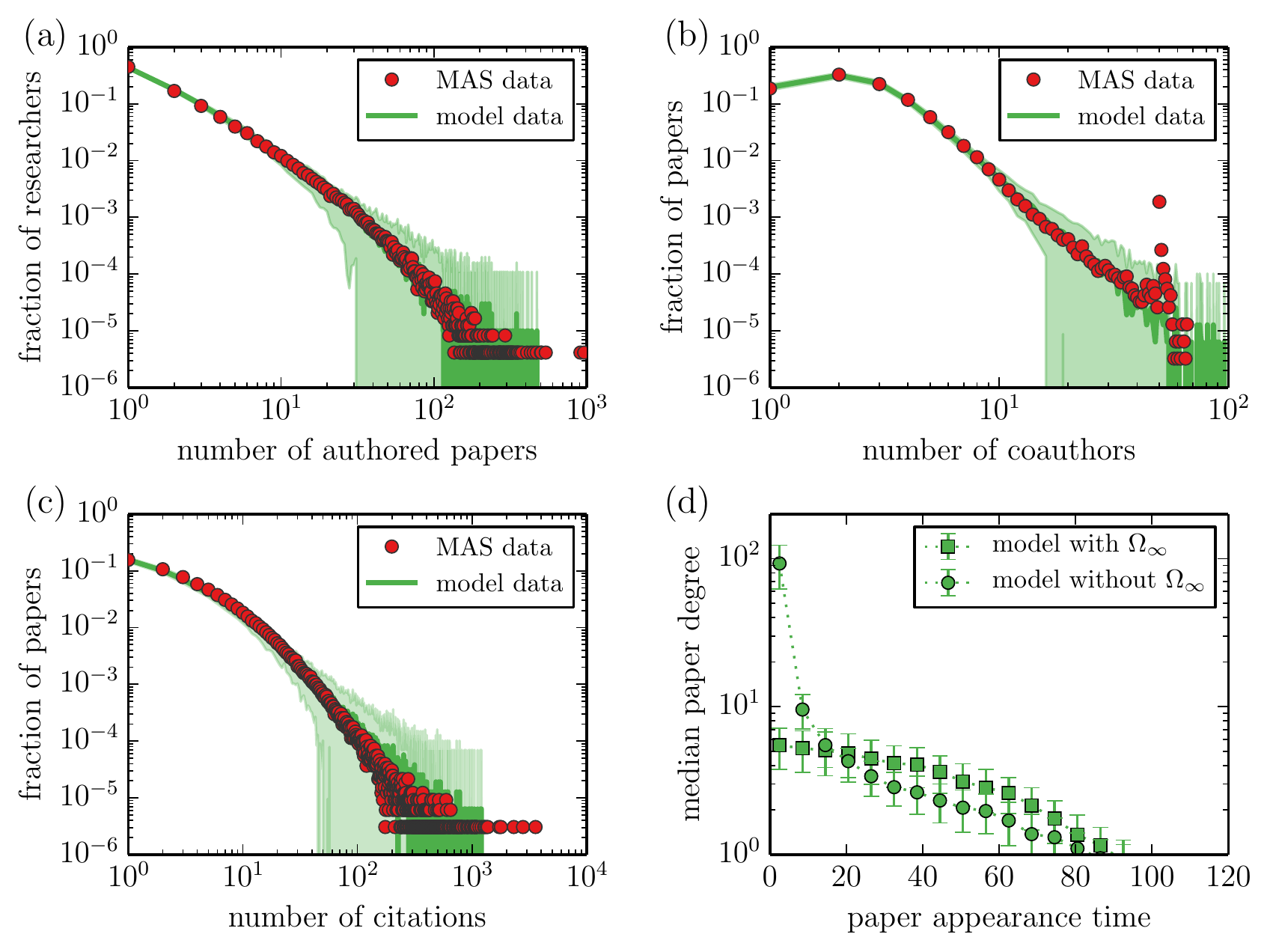}
\caption{Comparison of statistical features in the MAS data and in the artificial model: (a) distribution of researcher productivity $k$, (b) distribution of paper number of co-authors $d$, 
and (c) distribution of paper citation count $c$. In the log-log scale, the coefficients of determination ($R^2$) are 0.92, 0.85, and 0.85, respectively. 
The shaded areas visualize the variable's standard deviation observed in 100 model realizations. The dependence of paper citation count on paper appearance time 
with and without applying the stationary normalization value $\Omega_\infty$ is shown in panel~(d).}
\label{fig:calibration}
\end{figure*}

In each time step $t=1,\dots,T$, papers are gradually introduced in the system. For each paper $\alpha$, we build its set of authors $A_\alpha$ as follows. 
We first draw the number of authors $d_\alpha=\lvert A_\alpha\rvert$ from the distribution $\mathcal{H}(d)=19.7d/(100+d^{4.6})$, which is again motivated by the MAS data---see Figure~\ref{fig:empirical}(b). 
We then choose the $d_\alpha$ different authors, each with probability proportional to the remaining number of papers that they still have to author 
(for researcher $i$, this number is initially $k_i$, and then decreases by 1 with each authored paper). The fitness value $f_\alpha$ of the paper is proportional to the highest ability value among its authors: 
$f_\alpha=\eta \big(\max_{i\in A_\alpha}a_i\big)$ where $\eta$ is a multiplicative noise term that is uniformly distributed in $[1-\eta^*,1+\eta^*]$ and introduces additional randomness 
to the process of paper creation. Note that because of the extremal metric choice, paper fitness is not directly proportional to the number of authors. This assumption is motivated by recent empirical evidence: 
while papers with more authors receive on average more citations, this effect is apparently not related to papers quality~\cite{Bornmann2016}. 
Nevertheless, results obtained upon varying this and other assumptions are presented in the Supporting Information (SI).

Newly introduced papers make links to previously published papers. The probability that paper $\alpha$ cites paper $\beta$ at time $t$ is
\begin{equation}
\label{linkingP}
\mathcal{P}_{\alpha\to\beta}(t) = \frac{[c_{\beta}(t) + 1]\,f_{\beta}D(t - \tau_{\beta})}{\Omega(t)},
\end{equation}
where $c_{\beta}(t)$ is the current number of citations of paper $\beta$ and $\tau_{\beta}$ is its appearance time in the system, whereas, $D(\cdot)$ is the aging term and $\Omega(t)$ is the normalization term 
\begin{equation}
\label{Omega}
\Omega(t) = \sum_{\gamma}[c_{\gamma}(t) + 1]\,f_{\gamma}D(t - \tau_{\gamma}).
\end{equation}
Here, $c_{\beta}(t)$ needs to be increased by one to give a non-zero initial attractiveness to papers, as $c_{\beta}(\tau_\beta)=0$. In Equation~(\ref{linkingP}), we use exponential aging $D(t)=\exp(-t/\theta)$ 
where $\theta$ characterizes the lifetime of a paper. An alternative choice would be to use a log-normal shape for the aging term~\cite{Wang2013}.

Every new paper makes $q$ references to existing papers. Note that when a growing network with preferential attachment grows from a small initial configuration, 
papers that are present at early stages are in advantage with respect to later papers and can thus achieve a significantly higher citation count~\cite{Newman2009,berset2013effect}. 
Early papers enjoy the undue advantage during the initial period when $\Omega(t)$ is substantially smaller than its long-term stationary value $\Omega_\infty$. 
To overcome this problem, we assume that when $\Omega(t) < \Omega_{\infty}$, each of the $q$ links created by a newly added paper points to an existing paper with probability $\Omega(t) / \Omega_{\infty}$. 
In the complementary case, the link points out from the system and none of the existing papers receives it. This situation resembles a young scientific field which is growing, yet still partially relying 
on papers from other fields. The stationary value $\Omega_{\infty}$ is obtained by simulating the system for a sufficiently long time period and averaging the final $\Omega(t)$ 
over independent model realizations (values used in our simulations are specified in the SI). The complete simulation code can be found at \url{http://www.ddp.fmph.uniba.sk/~medo/physics/resources.html}.

\medskip

{\em Simulation parameters and dynamics}. We simulate systems with $A=1000$ authors over the time period of $T=120$ months. Paper lifetime is $\theta=48$ months, and $a_0$ is set to $1$ 
(note that while the use of $a_0$ is not essential, Figure \ref{fig:s1} in the SI shows that $a_0>0$ actually improves the agreement between empirical and model citation distribution). 
Each paper cites $q=20$ other papers, and fitness values are obtained with $\eta^*=0.2$. 
For our choice of the productivity distribution, the average author produces 5 papers and the most active author produces around 200 papers in total. 
In each month, several papers are introduced in the system so that, until the end of simulation, every author eventually produces the originally assigned number of papers.  
To achieve this, we endow each researcher with an activity counter $\nu_i(t)$, initially set to $k_i$. At step $t$ (when there are $T-t+1$ time steps left until the end of the simulation; $t=1,\dots,T$), 
we introduce new papers until the researcher activity counters decrease by $\sum_i \nu_i(t)/(T-t+1)$ in total. Due to the varying number of co-authors, the number of papers introduced at each step 
fluctuates but remains relatively stable during the whole simulation. The total number of papers produced in a single realization of the system is around 1500.

Figure~\ref{fig:calibration} reports basic calibration results for the model. In particular, panel (c) shows that the emergent citation distribution closely resembles the one observed for MAS data 
(see Figure \ref{fig:s1} in the SI for how the shape of this distribution changes when $a_0$ and $\theta$ are varied). Further, panel (d) shows that allowing some links to point out from the system 
indeed weakens the dependence of the paper citation count on its appearance time. For the present choice of parameters, correlation between paper citation count and fitness is around 0.5. In agreement 
with empirical studies of popularity in real systems~\cite{salganik2006experimental}, we see that while papers with high fitness on average attract more citations than papers with low fitness, 
there is still a substantial level of randomness in this relationship.

We conclude this section by remarking that we use empirical data from MAS only to calibrate the model: to measure the distributions of author productivity and of the number of co-authors per paper, 
and finally to fit the paper citation distribution. Since the shape of these distributions is rather general, using a different bibliometric dataset is not likely to qualitatively change the results of our analysis: 
the model is naturally flexible to adapt to other real datasets.

\section{Impact Indicators}
\label{sec:indicators}
We now introduce the indicators that we use to quantify the scientific impact of authors. In the following definitions, we will use quantities obtained at the end of simulations but omit the time label $T$. 
For instance, $c_\alpha$ denotes the number of citations paper $\alpha$ accrued at $t=T$. We define:

$\bullet$ {\em Total number of citations}, $C_i\equiv\sum_{\alpha:i\in A_\alpha}c_\alpha$.
The simplest possible choice, naturally favoring researchers with many papers and those who are active since long.

$\bullet$ {\em Average number of citations}, $E_i\equiv C_i/k_i$.
This approach is widely used in the literature, the underlying idea being that whenever a researcher receives a larger credit compared to the number of papers published, 
she is producing science having greater impact. Note that here we are considering only a single scientific field, and thus we do not need to worry about field-specific normalization~\cite{Radicchi2008}. 
In this way, average citations is equivalent to both the well-known $CPP/FCSm$ (citations per publication over mean field citation score) and $MNCS$ (mean normalized citation score) indicators~\citep{Waltman2011}.

$\bullet$ {\em Citation count of the most cited paper}, $M_i\equiv\max_{\alpha:i\in A_\alpha}c_\alpha$. 
This is an extremal metric that is influenced by the heavy-tailed distribution of the paper citation count, and thus should be used with caution.

$\bullet$ {\em $x$-index}~\cite{RodriguezNavarro2011,Bornmann2013}, the number of papers published by an author that are in the top $1\%$ most cited papers. This approach explicitly takes into account 
the extreme skewness of the citation distribution, which may cause average-based indicators to fail because of their sensitivity to the presence of one or a few very highly cited publications~\citep{Aksnes2004}. 
Percentile-based indicators like $x$ are instead less sensitive to these outliers~\citep{Waltman2013}.

$\bullet$ {\em $h$-index}~\cite{Hirsch2005}. Given the set ${\Pi}_i=\{\alpha_1,\dots,\alpha_\kappa,\dots,\alpha_{k_i}\}$ of papers authored by $i$ ordered by citation count in decreasing order 
(\ie, such that $c_{\alpha_\kappa}\ge c_{\alpha_{\kappa+1}}$, $\kappa\in[1,k_i-1]$), the $h$-index is the last position in which $c_{\alpha_\kappa}$ is greater than or equal to the position $\kappa$:
$$
h_i=\max_\kappa\left\{\min_{\alpha_\kappa\in \Pi_i}\left[c_{\alpha_\kappa},\kappa\right]\right\}.
$$

$\bullet$ {\em Contemporary $h$-index} ($hc$)~\cite{Sidiropoulos2007}, obtained by giving more weight to recent papers.
In particular, citations to papers published $\tau$ years ago are weighted with $4/(\tau+1)$. The $hc$-index is then computed as the $h$-index on the weighted citation counts.

$\bullet$ {\em $g$-index}~\cite{Egghe2006}.
Given the ordering ${\Pi}_i$, the $g$-index is the (unique) largest number such that the top $g$ articles received, together, at least $g^2$ citations:
$$g_i^2 \le \sum_{\substack{\kappa\le g_i \\ \alpha_\kappa\in{\Pi}_i}}c_{\alpha_k}.$$

$\bullet$ {\em $o$-index}~\cite{Dorogovtsev2015}.
Geometric mean of $M$ and $h$: $o_i=\sqrt{M_ih_i}$. The idea is that $M$ accounts for the researchers' greatest results and $h$ for their diligence. Thus, differently from the $h$-index, the $o$-index 
does not ignore the tail of the citation record.

$\bullet$ {\em Normalized $h$-index} ($n$, or {\em $m$-quotient})~\cite{Hirsch2005}, obtained as $m_i=h_i/\tau_i$ where $\tau_i$ is the time since the first publication of researcher $i$. 
This indicator is mainly aiming to identify young and promising scientists, as usually citation-based metrics favor senior researchers who had enough time to attract citations to their work \cite{vonBohlen2011}. 

\medskip

\begin{figure*}
\centering
\fig{0.7}{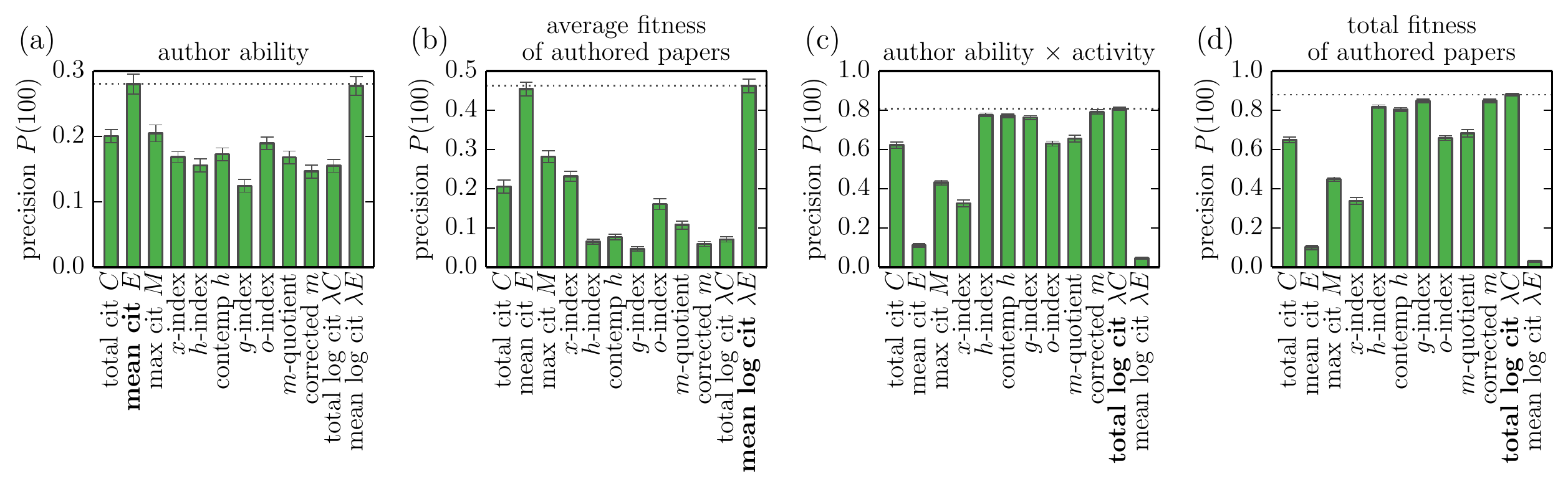}
\caption{Comparison of the mean precision values achieved by research impact indicators with respect to different ground truth assumptions. 
The horizontal dotted line marks the performance of the best metric (which is typed with bold letters); the error bars show three-fold of the standard error of the mean.}
\label{fig:results-basic}
\end{figure*}

Note that, apart from the $x$-index, all the metrics listed here can be computed for an individual researcher using solely her own citation records, \ie, without knowledge of citation statistics 
for the whole scientific community. This feature makes these indices apt for practical applications, as usually the whole dataset is unaccessible or very hard to handle (especially by individual researchers). 
In this respect, the inclusion of the $x$-index in our analysis is mainly for completeness---actually, $x$ is used mostly to compare not single researchers, but research institutions or communities at larger scales.

We remark that various metrics of scientific impact are not considered here for several reasons. For instance, in our simplified framework we model only one scientific area and one general publication venue, 
and thus it makes no sense to test metrics accounting for the research field \cite{Kaur2013} nor those based on comparing the total number of citations of a paper to those of other publications
in the same journal~\cite{Dodson2012,Smith2014}. Additionally, in our setting all co-authors are assumed to contribute to a paper equally, their activity decreases by 1 for every publication 
regardless of the number of co-authors, and the fitness of that paper does not scale with the number of co-authors. Thus, we cannot consider metrics that incorporate the relative contribution 
of each co-author to a paper, like the {\em individual h-index}~\cite{Batista2006} or the $SDC$ (``sequence-determines-credit'')~approach \cite{Teja2007} do. 
Nor we consider indices accounting for the quality of the citations in terms of the collaboration distance between citing and cited authors \cite{Bras2011}, 
as we do not model the presence of research groups. Finally, we do not consider metrics based on the {\em eigenvector centrality} within the citation network~\cite{Yan2009}, 
such as PageRank~\cite{walker2007ranking}, CiteRank~\cite{walker2007ranking} or PhysAuthorRank~\cite{radicchi2009diffusion}. This is because the linking probability defined by Eq.~(\ref{linkingP}) 
depends only on the fitness of the target paper $\beta$, and not on the fitness of the source paper $\alpha$. As a result, citations from a highly valued paper have the same intrinsic value 
as citations from an ordinary paper, and differentiating the weight of incoming citations thus cannot yield any improvements. Upon calibration on real data, our modeling framework can be extended 
to include many of the aforementioned effects and thus allow more metrics to be evaluated.

\section{Results}
We now assess the impact indicators against the ``ground truth'' provided by the intrinsic features of researchers in the model. In particular, we consider four different benchmarks: 
two intensive quantities (\ie, not depending on the number of authored papers), researcher ability $a_i$ and average fitness of authored papers $\overline{f_i} := \sum_{\alpha:i\in A_\alpha}f_\alpha/k_i$, 
and two extensive quantities (\ie, accounting for the number of authored papers), researcher ability times productivity $a_ik_i$ and total fitness of authored papers $\overline{f_i}k_i$. 
Note that for each pair of benchmarks, the first quantity refers to the researchers' potential, whereas, the second one is related to their actually realized publication output.

In order to evaluate a given impact indicator, we first use the ground truth to determine the set of 100 best authors. We then determine the set of the top 100 best-evaluated authors 
according to their impact score. Finally, we count the overlap $O$ between these two sets of researchers. The relative overlap $O/100$, which ranges from $0$ to $1$, is then a measure of the metric's performance 
(the higher the overlap, the more successful the impact metric in identifying the best researchers) which is commonly referred to as {\em precision} in information filtering literature~\cite{belkin1992information}. 
Note that in this specific setting, another classical metric, {\em recall}, is also equal to $O/100$~\cite{belkin1992information}. 
We evaluate precision achieved by individual metrics on 100 independent model evaluations, that are then used to compute the average precision and the standard error of the mean.

Figure~\ref{fig:results-basic} summarizes the metric evaluation results for the basic model setting described in Section~\ref{sec:model}. We see that when the ground truth is an intensive quantity 
(author ability in panel (a) or average fitness of the authored papers in panel (b)), the simple average citation score $E$ is by far the best performing indicator among the described group 
of traditional performance metrics. When considering extensive ground truths, the family of $h$-indices expectedly becomes more reliable: $h$, $hc$ and $g$ are tied for the first place. 
Since in this basic setting all researchers are in principle active from the beginning of the simulation, the $m$-quotient lags behind the original $h$ because of its uneven handling of researchers 
who have few papers and by chance started publishing late. The recently proposed $o$-index always performs midway between the $h$-index and $M$ (the citation count of the most cited paper), 
which suggests that the combination of these two quantities is not particularly effective in discerning the best researchers. The $x$-index does not perform well in any of the evaluations, 
mainly due to its reliance on a small subset of all papers (top 1\%) which makes it simultaneously a noisy and little discriminative metric for evaluation of individuals. 
Total citation count $C$ lags behind $h$-like metrics which is not surprising as it is highly sensitive to outliers.

To overcome the observed problems of certain metrics, we explore some variants that could possibly fare better. 
Firstly, to cope with outliers, we introduce the total logarithmic citation count $\lambda C_i=\sum_{\alpha:i\in A_\alpha}\log(c_\alpha + 1)$ 
(the citation count $c_{\alpha}$ is incremented by one to avoid $\log 0$ for papers with zero citations) and the mean logarithmic citation count $\lambda E_i = \lambda C_i / k_i$. 
Figure~\ref{fig:results-basic} shows that $\lambda E$ matches the good performance of the mean citation count $E$ for both intensive benchmarks. 
By contrast, $\lambda C$ slightly outperforms (approximately by 4\%) the established metrics for the extensive benchmarks, yet it has to be noted that the total logarithmic citation count 
is a considerably simpler metric than $h$ and $g$. This suggests that the use of a logarithmic unit of measure is an efficient way to deal with the skewness of the citation distribution. 
As for the $m$-quotient, its flaw is to allow many young authors who have only authored one or a few papers to score well, because their $h$-indices are divided by their small author age. 
While in real use this bias may be avoided by, \eg, selection committees enforcing explicit conditions on applicants (for instance, at least 3 years after the PhD defense, 
or a minimum number of publications), here we explore a mathematically grounded solution by formulating a {\em corrected $m$-quotient} ($cm$): $cm_i=(h_i/\tau_i)\times (1-1/\sqrt{k_i})$, 
where the second term penalizes researchers with very few authored papers (for example, $cm=0$ for all authors with only one publication).
As Figure~\ref{fig:results-basic} shows, such a corrected $m$-quotient then performs better than its original version and also outperforms the established metrics, though to a lesser extent than $\lambda C$.

Figure \ref{fig:s2} in the SI further shows how the performance of individual metrics change when individual model assumptions and parameters are varied. 
We see that while the choice of parameters has some impact on the achieved precision values, the best results are always obtained with the same set of metrics: $E$ and $\lambda E$ 
with respect to intensive ground truths, 
and the family of $h$-indices and $\lambda C$ with respect to extensive ground truths. Notably, the best performer $\lambda C$ is closely followed by the long-standing $h$-index in all studied settings, 
while some other well performing metrics slightly fall behind under certain circumstances (\eg, the $g$-index when the number of researchers in the simulated system is increased). 
Overall, we can conclude that the main results that we report here are robust with respect to substantial variations of the model and of its parameters.

\begin{figure*}
\centering
\fig{0.7}{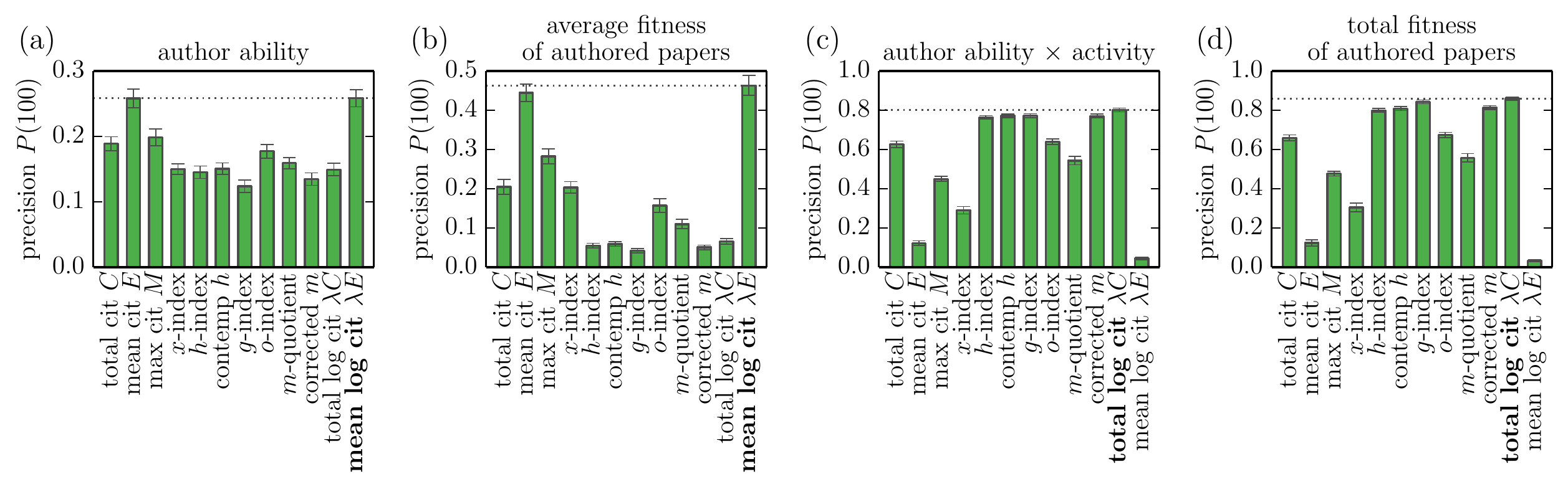}
\caption{Comparison of the mean precision values achieved by impact indicators with respect to different ground truth assumptions, when the number of active researchers grows in time (see the description in text). 
The horizontal dotted line marks the performance of the best metric (which is typed with bold letters); the error bars show three-fold of the standard error of the mean.}
\label{fig:results-changing_group}
\end{figure*}

We now wish to study the case where new authors are gradually introduced to the system, and check which indicators allow to fairly evaluate young researchers with respect to their senior colleagues. 
We thus consider a situation where $25\%$ of researchers are active for the whole time; for the remaining $75\%$, author $i$ begins their activity in a randomly chosen step $\tau_i$ between $t=1$ and $t=T-36$. 
Here 36 is subtracted to only include researchers who spent sufficient time in the system (by doing so, we are essentially considering only researchers who finished their PhD studies).
In this setting, the productivity of author $i$, initially drawn from $\mathcal{G}(k)$, is linearly rescaled by her appearance time in the system by a factor $(1-\tau_i/T)$: young authors are thus in disadvantage 
with respect to seniors by having on average less publications at the end of the simulation, and also by having had less time to accrue citations.

Figure~\ref{fig:results-changing_group} presents the results obtained with this setting. The first observation is that most metrics actually perform similarly than in the basic setting 
where the group of active researchers does not change over time. The only metric that considers authors' career length, $m$-quotient, is surprisingly performing worse in the new setting with researchers 
gradually entering the system. The reason lies again in the aforementioned rescaling problem: in the new setting, there are more young users who only author their papers in the last years 
and yet outperform venerable researchers upon the rescaling, thus lowering the resulting precision more than in the basic setting presented in Figure~\ref{fig:results-basic}. 
Specifically, there are on average $28\pm8$ authors with only one publication in the top 100 positions of the ranking by the $m$-quotient and the average activity span of top 100 researchers is $60\pm 40$ months 
(out of $120$ in total). By contrast, for the $h$-index ranking there are no authors with only one publication in top 100 and the average activity span of top researchers is $96\pm 21$ months. 
The bias towards very young researchers is removed by the corrected $m$, which brings to no researchers with only one paper in top 100 and to an average activity span of $86\pm28$ months: 
the resulting precision is similar to that achieved with the $h$-index and the contemporary $h$-index. Overall, the logarithm-based indices $\lambda E$ and $\lambda C$ are again the best performers 
against intensive and extensive ground truths, respectively.

We conclude by discussing the choice of optimal benchmarks to evaluate impact indicators against. On one hand, extensive ground truths are more appropriate when we assume that all researchers are active 
for the whole time period. This is because intensive benchmarks neglect productivity, and thus do not penalize gifted but little active researchers who produce only a few (though very good) papers. 
On the other hand, when considering researchers with different activity lifetime, extensive ground truths automatically give preference to authors who are active since longer, 
thus using intensive benchmarks may be more appropriate in this case. Yet, because of their own nature, intensive criteria are unable to properly handle authors 
with very low activity and thus little citation statistics. Relying on an ``intermediate'' ground truth could represent a suitable solution, albeit its precise form is certainly arbitrary. 
Our artificial framework makes it easy to evaluate impact indicators with respect to a different ground truth assumption. Figures~\ref{fig:s3} and \ref{fig:s4} in the SI show results for two intermediate benchmarks: 
researcher ability times square root of productivity $a_i\sqrt{k_i}$ and an analogous multiple of the average fitness of the authored papers $\overline{f_i}\sqrt{k_i}$. 
We observe no major changes with respect to the purely extensive benchmarks, supposedly because activity values are broadly distributed: even after the square root is applied, 
substantial activity is still needed to access the top 100 of the new ground truth. From the viewpoint of evaluating researchers, the dual intensive-extensive approach used here thus seems sufficient. 
There are some particular aspects though, such as the ability of a metric to identify young talented researchers, that can only be captured by ground truths that specifically target at the feature of interest 
(researchers active for less than six years, for example). Construction of such ground truths and their use in the proposed model-based evaluation framework remain as open issues for future research.

\section{Discussion}
The ongoing proliferation of scientific impact indicators is facilitated by the critical lack of solid evaluation criteria. 
Presently, motivation for new metrics is sometimes only anecdotal and their evaluation often relies on outliers analysis~\cite{Hirsch2005,Egghe2006,Sidiropoulos2007,radicchi2009diffusion,Dorogovtsev2015}. 
However, outliers in any metric are almost inevitably highly successful authors or highly cited papers---such validation is thus very soft and eliminates only the most ill-suited metrics. 
The absence of a ``golden standard'' (certified best papers and researchers) for validation of indicators on real data compels the use of various {\em ad-hoc} proxies, 
such as relying on experts' judgment \cite{,Bornmann2016}. In this work, building on an artificial model of citation dynamics, we have established a test bench where new and old metrics can face their first examination. 
The use of a controlled framework allowed us to avoid the biases present in real citation databases related to coverage issues and to improper citation practices~\cite{Jacso2006,Wildgaard2014,Werner2015} 
and, more importantly, to have ground truth features to evaluate impact indicators against in a quantitative way.

Our framework, which generates bibliometric statistics whose aggregate characteristics closely match those of real citation data (Figure~\ref{fig:calibration}), 
is based on a number of assumptions and simplifications, yet it is open to include additional features of real citation dynamics. 
For instance, we could consider several research areas with different citation rates, which would in turn allow us to study field normalization for impact indicators.
Additionally, while we have intentionally excluded journals, review process, and the impact of publishing venue on paper success, different journals could be included to model important aspects 
for the dynamics of paper popularity, like high impact factor journals having a broader readership and attracting more citations for their articles. 
Moving further, the social network of researchers plays an important role in real citation dynamics, and thus in principle we could consider the presence of both befriended and competing scientists, 
the structure of research collaborations~\cite{borner2004simultaneous}, and the feedback of author reputation on the dynamics of paper popularity~\cite{Petersen2014}. A strong assumption of our model 
is the use of Eq.~(\ref{linkingP}) that determines the citation mechanism. Indeed, while this formulation was shown to fit real data better than any other model proposed so far~\cite{Medo2014}, 
it can of course be improved. For instance, we could vary the sensitivity of citing papers to cited papers quality, \ie, make high fitness papers more likely to cite other good papers than low fitness papers do. 
Besides making the model more realistic (there is empirical evidence that highly cited papers do cite other highly cited papers more often than one would expect, in particular more often 
than the badly cited papers do~\cite{Bornmann2010}), such a modification could allow PageRank-like metrics to yield results superior to simple local metrics such as citation count and $h$-index.

Despite these simplifications, our analysis allowed us to quantitatively asses what impact indicators actually reflect. We found that the average citation score efficiently measures authors ability, 
whereas, the $h$ and $g$ indices and the simpler cumulative logarithmic citation count $\lambda C$ do capture joint ability and productivity of researchers. 
Additionally, we provided several recommendations for a proper use of the normalized $h$-index in the identification of talented young scientists.
While our results are only preliminary, and may become more robust by equipping the model with more realistic assumptions, 
we remark that the issue of studying what impact indicators do measure is of critical importance nowadays, as these metrics are currently so widely employed by selection committees 
that basically determine ``most things that matter: tenure or unemployment, a postdoctoral grant or none, success or failure''~\cite{Lawrence2008}. 
However, impact indicators are ``usually well intentioned, not always well informed, often ill applied''~\cite{Hicks2015}. In other words, while these indicators have been designed to improve the system, 
their improper use is putting the system in danger---primarily by modifying the very aim of scientists from making discoveries to publishing as many papers and getting as many citations as possible. 
In this respect, simulation scenarios may ease the difficulties in determining what a given measurements of scientific impact reflects, 
without overlooking the fact that in any case impact indicators alone cannot be used to judge individual scientists.

\begin{acknowledgments}
We acknowledge support from the EU projects GROWTHCOM (FP7-ICT, no. 611272), MULTIPLEX (FP7-ICT, no. 317532) and CoeGSS (EINFRA, no. 676547).
The funders had no role in study design, data collection and analysis, decision to publish, or preparation of the manuscript. 
We thank the group of Professor Frank Schweitzer from ETH Zurich for providing us with the aggregate results for MAS data that are presented in Figures~\ref{fig:empirical} and \ref{fig:calibration}.
\end{acknowledgments}

\setcounter{table}{0}
\setcounter{figure}{0}
\renewcommand{\thetable}{S\arabic{table}}
\renewcommand{\thefigure}{S\arabic{figure}}
\renewcommand{\theHtable}{Supplement.\thetable}
\renewcommand{\theHfigure}{Supplement.\thefigure}

\onecolumngrid
\vspace{2cm}

\section*{Supporting Information}

\begin{figure*}[h!]
\centering
\fig{0.7}{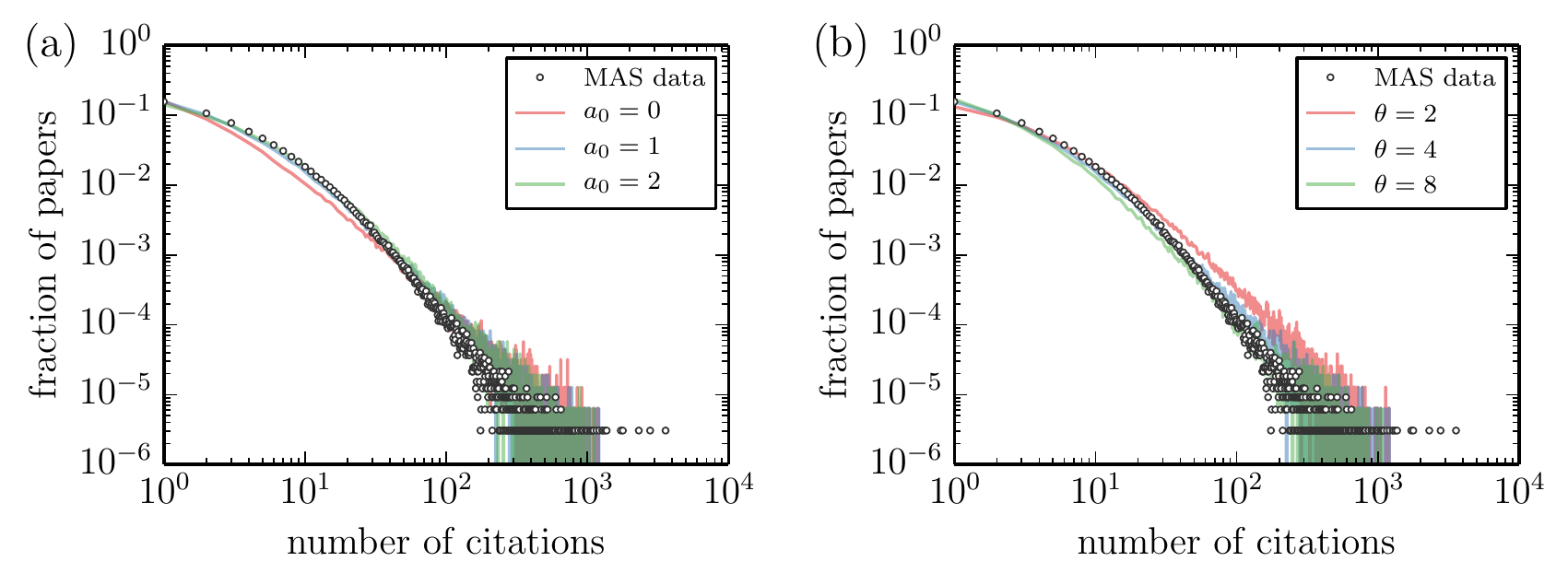}
\caption{The effect of parameters variations on the paper citation distribution: (a) baseline author ability $a_0$ (when $\theta=4$), and (b) paper lifetime $\theta$ (when $a_0=1$). 
The left panel shows that introducing a value $a_0>0$ helps to make the distribution more concave. 
The right panel shows that tuning paper lifetime can be used to calibrate the simulated citation distribution against the real one: 
in the log-log scale, the coefficient of determination ($R^2$) increases from $0.60$ for $\theta=2$ years to $0.85$ for $\theta=4$ years. 
The stationary value for the normalization constant $\Omega_{\infty}$ is $1.33\cdot 10^4$ in the basic setting $a_0=1$ and $\theta=4$. 
For the variations considered here, $\Omega_{\infty}$ values are $1.06\cdot 10^4$ and $1.62\cdot 10^4$ for $a_0=0$ and $a_0=2$, respectively, 
and $6.30\cdot 10^3$ and $2.17\cdot 10^4$ for $\theta=2$ and $\theta=8$, respectively.}
\label{fig:s1}
\end{figure*}


\begin{figure*}
\centering
\fig{0.7}{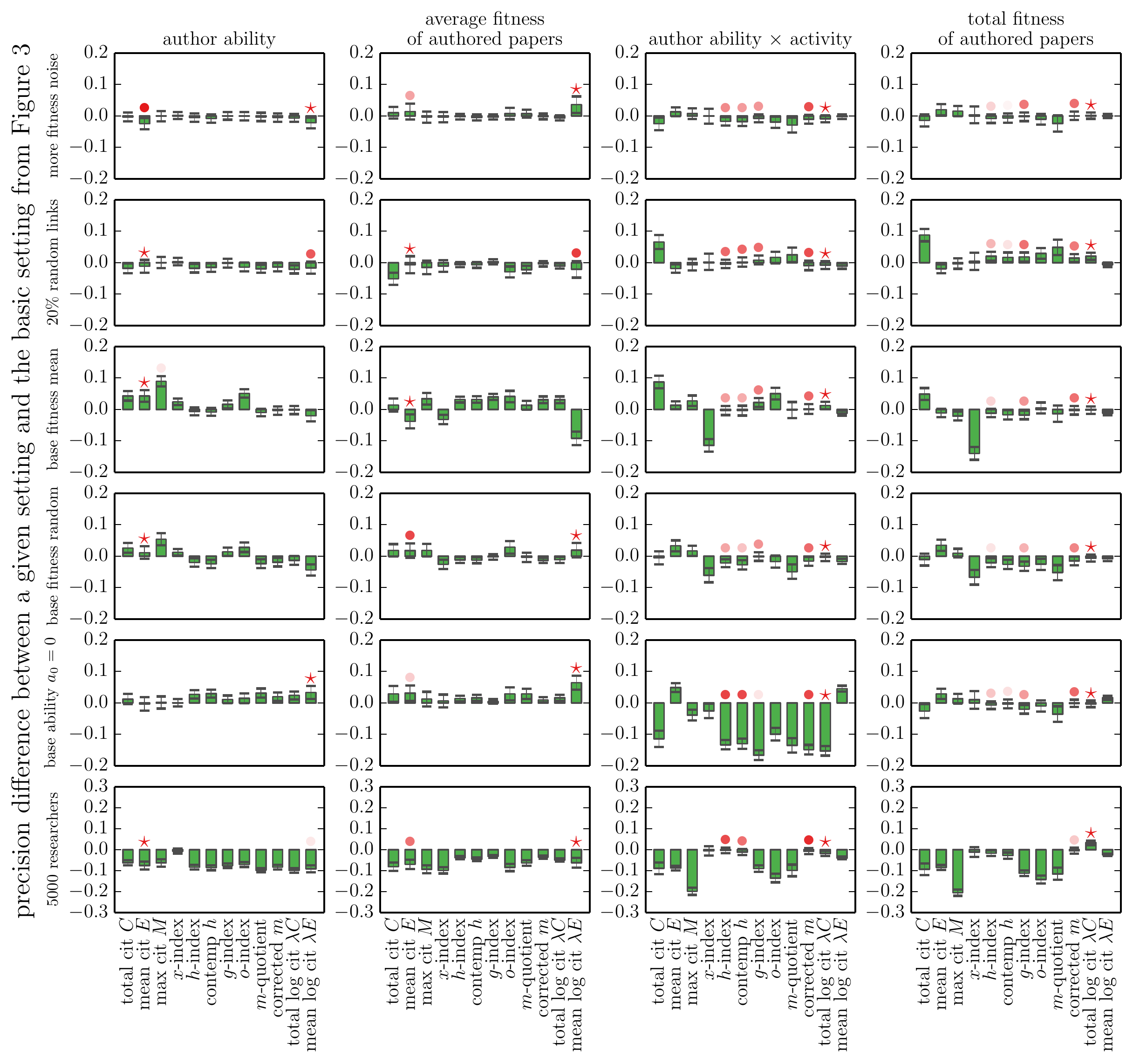}
\vspace*{-12pt}
\caption{Difference in precision---with respect to the basic setting---for individual metrics when various variations of this setting are considered (from top to bottom):\\
(1) the fitness noise amplitude $\eta^*$ is increased from 0.2 to 0.5 ($\Omega_{\infty}=1.48\cdot 10^4$),\\
(2) 20\% of references made by a new paper target a random existing paper ($\Omega_{\infty}=1.28\cdot 10^4$),\\
(3) the base fitness of a paper is obtained as the average of the authors' ability values ($\Omega_{\infty}=9.20\cdot 10^3$),\\
(4) the base fitness of a paper is a random sample from the authors' ability values ($\Omega_{\infty}=1.10\cdot 10^4$),\\
(5) the baseline ability of authors $a_0$ is set to zero ($\Omega_{\infty}=1.06\cdot 10^4$),\\
(6) the number of researchers is increased from 1000 to 5000 ($\Omega_{\infty}=6.90\cdot 10^4$).\\
The error bars show the three-fold of the standard error of the mean. For every setting and ground truth, the best-performing metric (in absolute terms) is marked with a star ($\star$), 
and all metrics that reach at least 90\% of its precision are marked with bullets ($\bullet$) whose color indicates the performance difference 
(\textcolor{r_dark}{$\bullet$}, \textcolor{r_middle}{$\bullet$}, and \textcolor{r_pale}{$\bullet$} correspond to 98\%, 95\%, and 92\% of the best performance, respectively).}
\label{fig:s2}
\end{figure*}


\begin{figure*}
\centering
\fig{0.7}{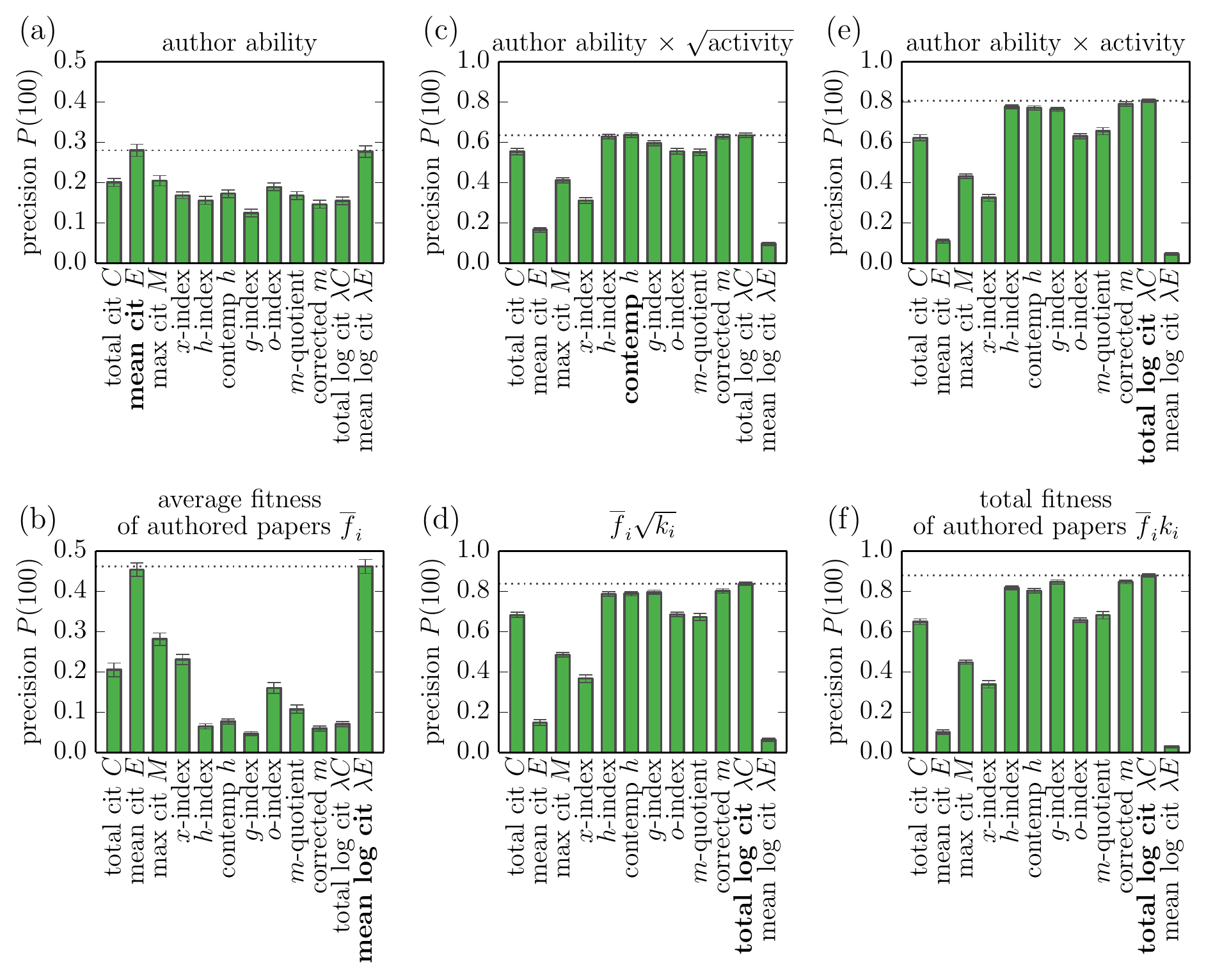}
\caption{Precision values achieved by individual metrics against intensive (a,b), midway (c,d) and extensive (e,f) ground truth assumptions 
(we assume here the basic simulation setting that was used to obtain Figure 3 in the main text).  The first row of panels (a,c,e) refers to a benchmark obtained from the researchers' potential, 
whereas, the second row (b,d,f) to a benchmark related to realized publication outputs. Here $\Omega_{\infty}=1.33\cdot 10^4$.}
\label{fig:s3}
\end{figure*}


\begin{figure*}
\centering
\fig{0.7}{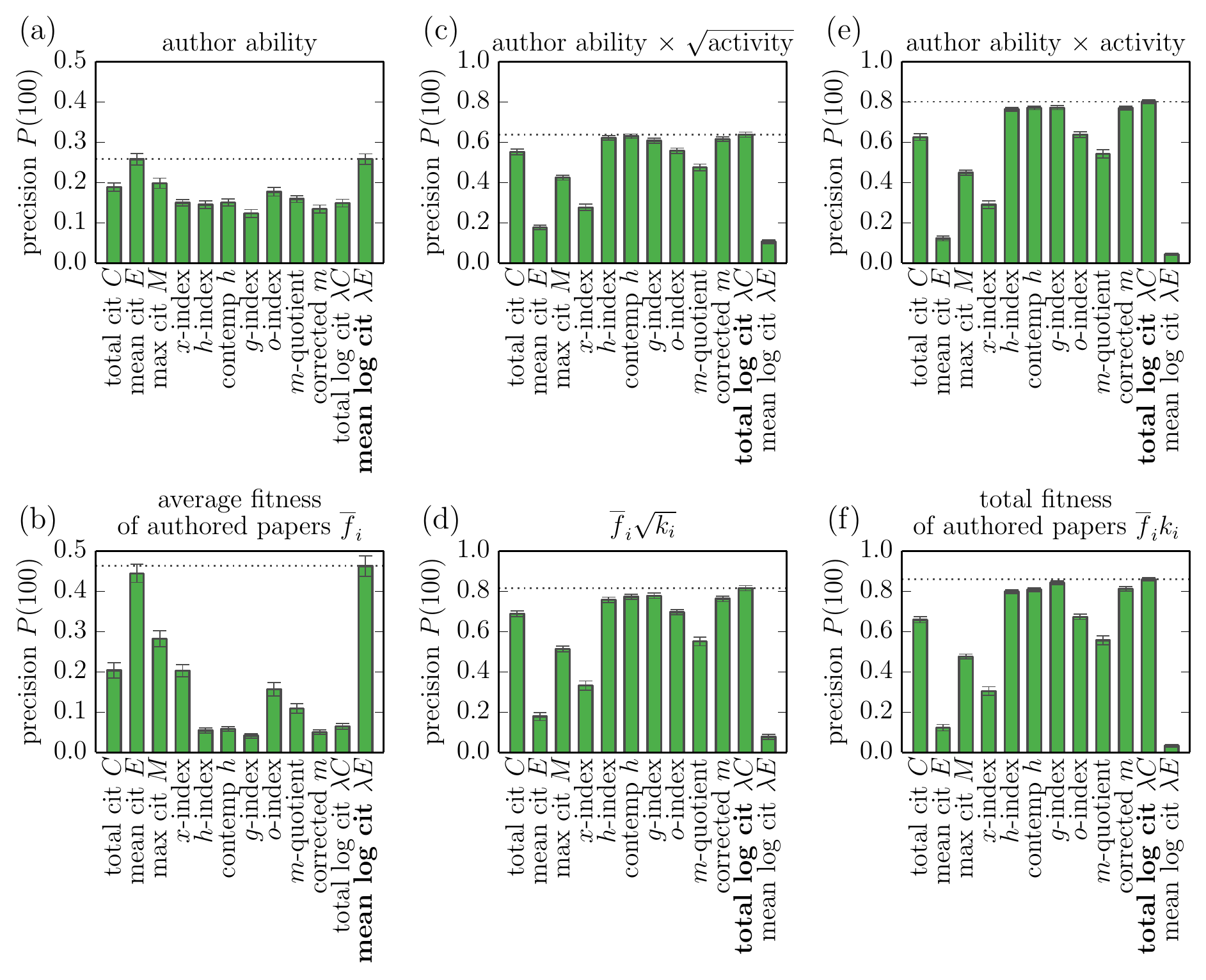}
\caption{Same as FIG. S3 but for the simulation setting with the number of active researchers increasing with time as in Figure 4 in the main text. Here $\Omega_{\infty}=1.10\cdot 10^4$.}
\label{fig:s4}
\end{figure*}

\end{document}